# Observation of Spin Hall Effect in Weyl Semimetal WTe$_2$ at Room Temperature

Bing Zhao[1,2], Dmitrii Khokhriakov[2], Yang Zhang[3], Huixia Fu[4], Bogdan Karpiak[2], Anamul Md. Hoque[2], Xiaoguang Xu[1], Yong Jiang[1], Binghai Yan[4], Saroj P. Dash[2,5,*]

[1]*Beijing Advanced Innovation Center for Materials Genome Engineering, School of Materials Science and Engineering, University of Science and Technology Beijing, Beijing 100083, China*
[2]*Department of Microtechnology and Nanoscience, Chalmers University of Technology, SE-41296, Göteborg, Sweden*
[3]*Max Planck Institute for Chemical Physics of Solids, 01187 Dresden, Germany*
[4]*Department of Condensed Matter Physics, Weizmann Institute of Science, Rehovot 7610001, Israel*
[5]*Graphene Center, Chalmers University of Technology, SE-41296 Göteborg, Sweden.*

**Abstract**

**Discovery of topological Weyl semimetals has revealed the opportunities to realize several extraordinary physical phenomena in condensed matter physics. Specifically, these semimetals with strong spin-orbit coupling, broken inversion symmetry and novel spin texture are predicted to exhibit a large spin Hall effect that can efficiently convert the charge current to a spin current. Here we report the direct experimental observation of a large spin Hall and inverse spin Hall effects in Weyl semimetal WTe$_2$ at room temperature obeying Onsager reciprocity relation. We demonstrate the detection of the pure spin current generated by spin Hall phenomenon in WTe$_2$ by making van der Waals heterostructures with graphene, taking advantage of its long spin coherence length and spin transmission at the heterostructure interface. These experimental findings well supported by ab initio calculations show a large charge-spin conversion efficiency in WTe$_2$; which can pave the way for utilization of spin-orbit induced phenomena in spintronic memory and logic circuit architectures.**

**Main**

A strong resurgence of interest in two-dimensional (2D) transition metal dichalcogenide (TMD) is sparked with the successful preparation of materials with different properties that have the potential to revolutionize the future of electronics[1,2]. While semiconducting TMDs brings enormous interest in transistors[3–6] and optoelectronic applications[7]; the semimetals are predicted to host novel topological electronic states[8–10]. The recently predicted type-II Weyl semimetals[10–12] such as $WTe_2$ shows extraordinary electronic phenomena, like – a giant magnetoresistance[13], high mobilities[14], chiral anomaly[15] and anomalous Hall effect[16]. These novel transport features indicate the existence of Weyl fermionic states, which are characterized by a tilted linear dispersion of Weyl cones and Fermi arc surface states. Due to the monopole-like Berry curvature in the momentum space, strong spin-orbit interaction, a unique spin texture in Weyl cones and Fermi arc surface states are predicted to exist[17–19]. In addition to the topological Weyl features in these semimetals, trivial spin-polarized Fermi arc surface states are also shown to exists at the Fermi level between the electron and the hole pockets at room temperature[20–26]. Taking advantage of these properties, recent experiments with $WTe_2$/ferromagnet bilayers showed a control of spin-orbit torque arising from its crystal symmetry[27,28]. Therefore, these 2D semimetals are considered to have a huge potential for ultra-low power spintronic devices[29] with an efficient conversion of charge-to-spin current, i.e. a large spin Hall effect (SHE) and (or) Rashba Edelstein effect (REE) at room temperature[30], however, it has not been yet experimentally measured.

Here, we report an observation of a large spin Hall effect (SHE) in semimetal $WTe_2$ devices at room temperature. We electrically detect the SHE signals by employing a van der Waals heterostructure device of $WTe_2$ and graphene, taking advantage of 2D layered structures of both classes of the materials. In these experiments, we exploit the best of both the worlds, such as a large spin Hall angle of $WTe_2$, along with a long spin coherence length in graphene and an efficient spin transfer at the $WTe_2$-graphene interface. The large charge-spin conversion signal stems mainly from bulk SHE of $WTe_2$ and possibly REE from the $WT_2$ surface states. Our detailed spin sensitive electronic measurements both in the in-plane and perpendicular geometries, its angle and gate dependent studies, and theoretical calculations manifest the existence of the large spin Hall phenomena in $WTe_2$ devices at room temperature.

**Results and Discussions**

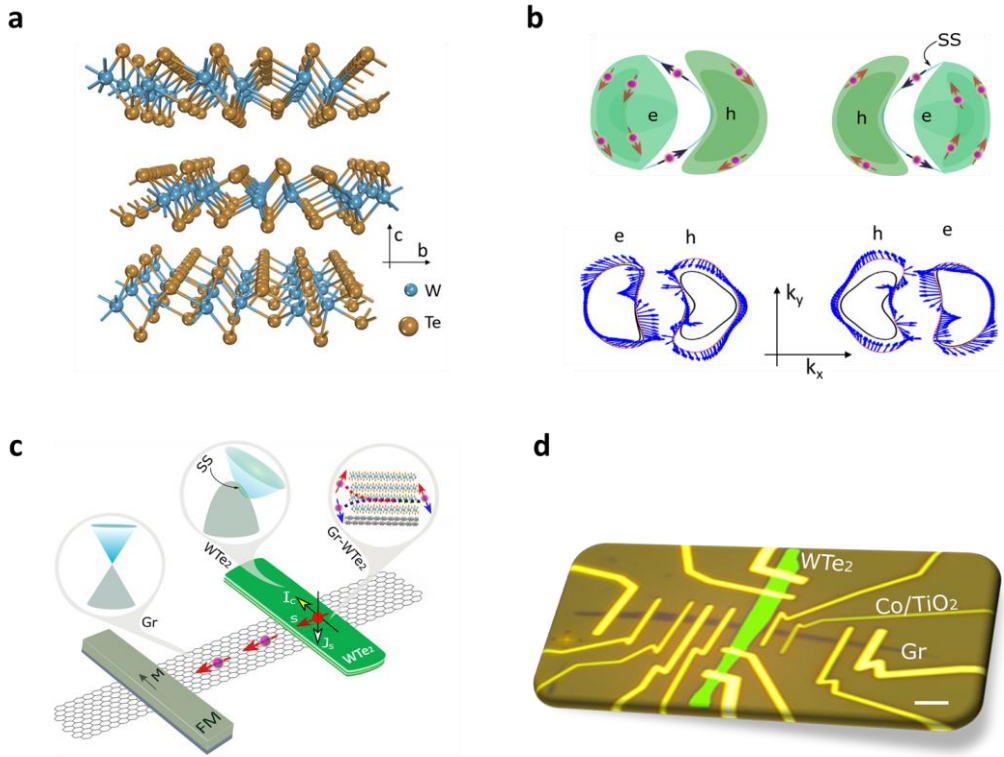

*Figure 1. Scheme for detection of spin Hall effect and inverse spin Hall effect in WTe$_2$. **a.** Crystal structure of the WTe$_2$ in T$_d$ phase showing the layered nature. **b.** Top: schematic of the spin textures at the Fermi surface with the electron (e), hole (h) pockets and the surface states (SS). Bottom: the calculated electronic structure showing the coexistence of both the e and h pockets on the Fermi surface. Both the electron and hole bands split into two bands (red and black). The spin texture is indicated by blue arrows in one of the split bands (red). The two split bands have opposite directions of spin polarization. **c,d.** Schematics and colored optical microscope picture of a nanofabricated WTe$_2$–graphene van der Waals heterostructure device with WTe$_2$ flake (green), ferromagnetic tunnel contacts of TiO$_2$/Co (FM) on the graphene (Gr) channel for the measurement of SHE and the ISHE on a SiO$_2$/Si substrate. The insets in the schematics show the band structures of the materials and the structure at the interface. The scale bar (white) in the device picture is 2μm.*

Figures 1a and 1b show the crystal structure and electronic structure, respectively, for WTe$_2$ in the T$_d$ phase. At the Fermi surface, small electron and hole pockets coexist, demonstrating a compensating semi-metallic feature. Because of the inversion symmetry breaking and strong spin-orbit coupling (SOC), each pocket splits into two bands (see Fig. 1b). The Fermi surface exhibits a clear spin texture. The left and right parts of the Fermi surface and spin texture can be transformed into each other by a mirror reflection ($k_x$ to $-k_x$). Such as strong spin-momentum locking feature indicates that the charge current comes together with a spin current, like SHE and REE. We experimentally investigated the influence of the spin degree of freedom on the charge currents and vice versa due to the presence of strong SOC, broken inversion symmetry and the novel spin textures in WTe$_2$. The SHE in WTe$_2$ is expected to cause a transverse spin current induced by a charge current, whereas, the inverse SHE (ISHE) produces a transverse charge current that is caused by a pure spin current[31,32]. Figure 1c,d show the nanofabricated devices consisting of van der Waals heterostructures of $WTe_2$ with few-layer graphene having ferromagnetic tunnel contacts to detect the SHE and ISHE in a spin sensitive potentiometric measurement (see Methods for details about the fabrication process). The heterostructure of graphene with $WTe_2$ flakes of 11-30 nm in thickness was used (from Hq Graphene), as measured by the atomic force microscope (AFM) (Supplementary Fig. S1). The quality of the $WTe_2$ was characterized by Raman spectrometer, showing peaks corresponding to the $T_d$-phase (Supplementary Fig. S2).

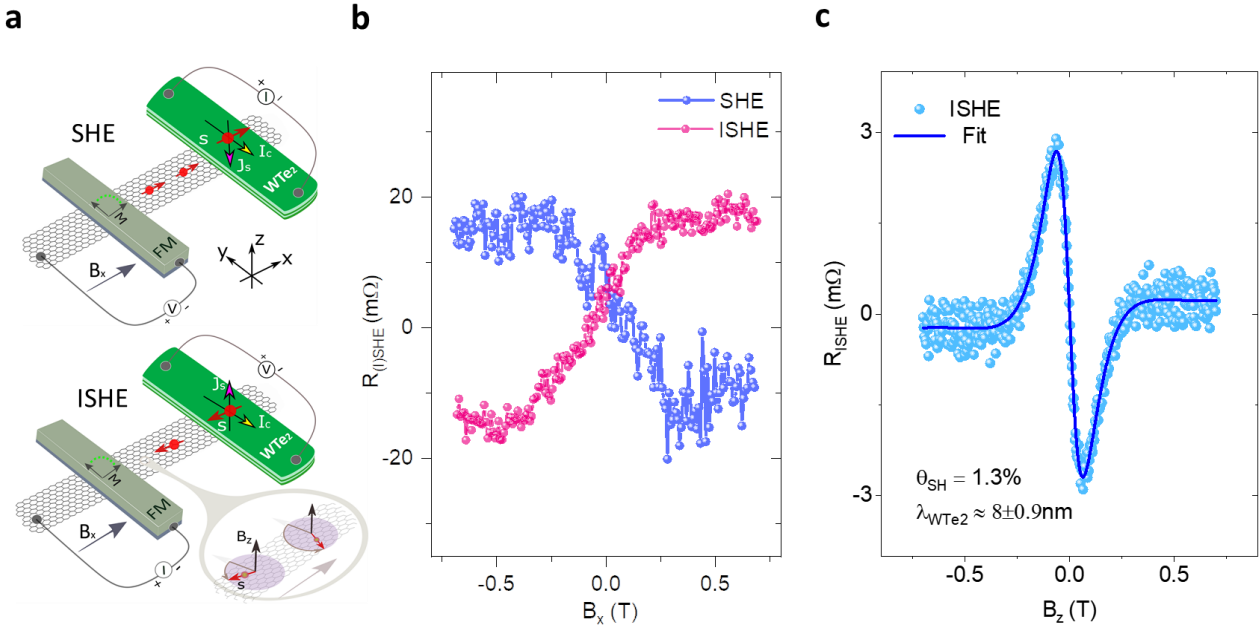

*Figure 2. Electrical detection of spin Hall effect and Inverse spin Hall effect in $WTe_2$ at room temperature. a. Top: Schematic diagram of SHE measurement configuration, where a charge current in $WTe_2$ induces a spin current, which is injected and detected in graphene using a ferromagnetic contact in a non-local geometry, while the direction of the applied magnetic field sweep is $B_x$. Bottom: Schematic representation of ISHE measurement configuration, where a spin current injected from a ferromagnetic electrode into graphene enters the $WTe_2$ channel and hence induces a charge current giving rise to a voltage signal in $WTe_2$, while the applied magnetic field sweeps along $B_x$ and $B_z$, respectively. Inset is the out-of-plane $B_z$ induced the spin precession. b. The change in SHE and ISHE resistance ($R_{(I)SHE}=V_{(I)SHE}/I$) with magnetic field $B_x$ sweep due to the (I)SHE for bias currents of +60 µA at room temperature for Dev 1 with channel length $L_{SHE}$=2.6 µm between Co and $WTe_2$. c. ISHE signal data and fitting curve with out-of-plane magnetic field $B_z$ in Dev 3 with channel length $L_{SHE}$=3.5 µm at bias currents of +100 µA. The ISHE signal data shown here were an average of 5 repeating measurement field sweeps. All the measurements were performed at 300K.*

Figure 2a show the measurement geometries for SHE and ISHE in a hybrid device consisting of $WTe_2$-graphene heterostructure and a ferromagnetic contact. The contact resistance of $WTe_2$-graphene in the Dev 1 during SHE measurements was around ~ 2kΩ•µm$^2$ (Supplementary Fig. S3a). The application of longitudinal charge current (*I*) in $WTe_2$ produces a pure transverse spin current due to SHE, which is injected into the graphene at the interface and subsequently detected as a voltage signal ($V_{SHE}$) by the non-local ferromagnetic Co tunnel contacts. The direction of the injected spins is in the plane of the graphene and perpendicular to the ferromagnet electrodes. The magnetic field $B_x$ is applied perpendicular to the electrodes for changing the magnetization direction of Co from 90° to 0° with respect to the injected spins. Figure 2b shows the measured SHE data of $R_{SHE}=V_{SHE}/I$ for I=60 µA at room temperature for a $B_x$ sweep for Dev 1 with graphene channel length $L_{SHE}$=2.6µm. As expected, $R_{SHE}$ follows a linear dependence at low B field, due to sin(θ) dependence of the ferromagnetic moments rotation angle θ with the Co electrodes, whereas at large enough B field, the magnetization of Co rotates 90° and become parallel to B field and also the injected spin directions, resulting in the saturation of $R_{SHE}$.

Next, we performed the ISHE experiment, where a pure spin current is injected from the ferromagnet and absorbed by the $WTe_2$. The spin current at the $WTe_2$-graphene interface should give rise to a transversal charge voltage ($V_{ISHE}$) due to the ISHE (Fig. 2a). Figure 2b shows the measured ISHE data of $R_{ISHE}=V_{ISHE}/I$ for I=60 µA at room temperature with sin(θ) behavior for a $B_x$ field sweep. These observed features confirm that the measured signal arises from spin to charge conversion in $WTe_2$. According to our measurement geometry and the SHE signal (Fig.2a and 2b), the spin hall angle $\theta_{SH}$ is positive based on $I_s \propto s \times I_c$[33]. This is confirmed by a bias current polarity dependence of the (I)SHE signals (Supplementary Fig. S5). Both the signals $R_{SHE}$ and $R_{ISHE}$ saturate with the magnetization of the injector/detector ferromagnetic Co electrode, as verified from the spin precession Hanle measurements with $B_x$ field in graphene channels (see Supplementary Fig. S7). The observed comparable SHE and ISHE signal magnitudes, their line shapes with magnetic field sweeps are in agreement with the Onsager reciprocity relation[34] and demonstrate the generation and detection of pure spin currents in $WTe_2$.

To further verify the charge-spin conversion, out-of-plane $B_z$ field sweep measurements were also performed in the ISHE configuration in Dev 3. This Dev 3 consists of monolayer graphene, 11 nm $WTe_2$ with 1 µm width and 25 Ω·µm² graphene-$WTe_2$ interface resistance (see Supplementary Table 1 for details about the device). The spin current injected from the FM electrode experience a spin precession in the graphene channel (L=3.5µm) as the $B_z$ field is perpendicular to the graphene plane. Subsequently, the spin current gets absorbed ~100 % at the graphene/$WTe_2$ interface for the monolayer graphene device used here (see Supplementary Fig. S9 d) and give rise to a transversal charge voltage ($V_{ISHE}$), where $R_{ISHE}=V_{ISHE}/I$ (Fig. 2c). Contrary to the in-plane ISHE with $B_x$ field sweep, here the magnetization of FM does not rotate with the $B_z$ field in this field sweep range and remains in-plane. The observed $R_{ISHE}$ is antisymmetric with $B_z$ field. At $B_z$=0, the injected spins are oriented along with the $WTe_2$ flake without any precession and results in $R_{ISHE}$=0 as it lacks the right-hand rule for the observation of ISHE[33,35] in the measured geometry. At a finite $B_z$ field, a maximum (minimum) values of $R_{ISHE}$ are obtained when the precession provides a spin component perpendicular to the $WTe_2$ long axis. Finally, at a larger $B_z$ field, $R_{ISHE}$ decreases with an increase in spin precession angle and approach to zero due to complete spin dephasing[36]. The in-plane ($B_x$) ISHE measurements in the same Dev 3 is presented in Supplementary Figure S6. Both the in-plane ($B_x$) and out-of-plane ($B_z$) measurements unambiguously demonstrate that the in-plane spins are responsible for the induction of the ISHE signal in $WTe_2$.

The magnitude of the measured (I)SHE signals (up to ~30 mΩ) in $WTe_2$ are very large, which are two orders of magnitude larger than measured in metals (Pt) in heterostructures with Cu[37], and 3 times larger than Pt in heterostructures with graphene[36,38], indicating a very large spin Hall angle $\theta_{SH}$ in $WTe_2$. This can be confirmed by the large spin valve signal reduction since the spin absorption by $WTe_2$ is the precondition to observe the ISHE[38,39] (see details in supplementary information Note 1).

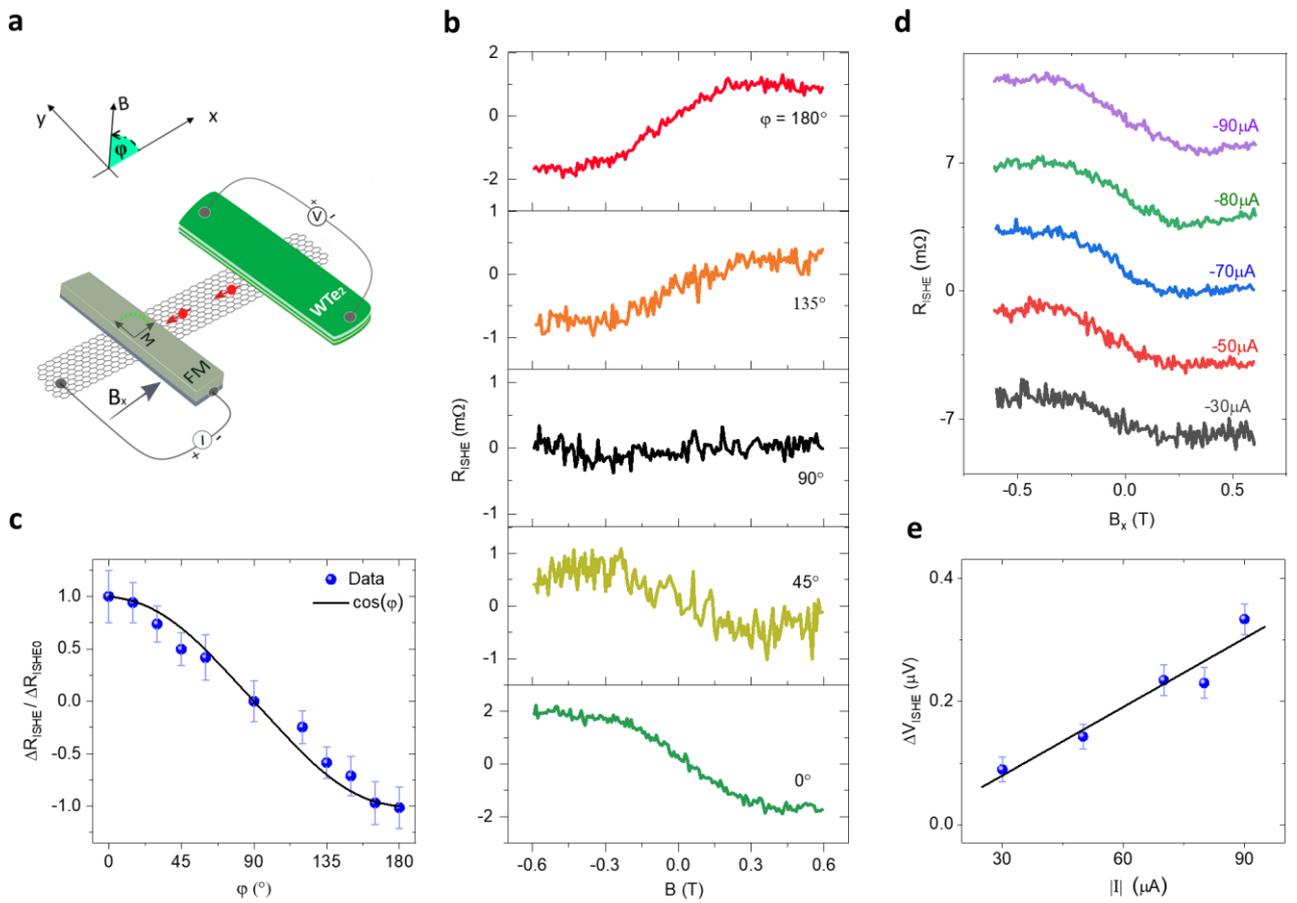

*Figure 3. Angle dependence of Inverse spin Hall effect in $WTe_2$. **a.** Schematic representation of the ISHE measurement geometry with directions of an applied magnetic field, the detector ferromagnet magnetization, and the spin current. The angle φ is defined as shown in the inset. **b.** The ISHE resistance $R_{ISHE}$ measured at room temperature for various measurement angle orientations for Dev 2 with a graphene channel length of $L_{SHE}$ = 3.5 μm. **c.** The normalized magnitude of $\Delta R_{ISHE}$ as a function of the magnetic field angle φ. The solid line is the cos(φ) curve. **d,e.** The measured $R_{SHE}$ at different spin injection bias currents with a shift at y-axis for the sake of clarity and the magnitude of the ISHE signal with an applied bias current for the Dev 2. All the measurements were performed at 300K.*

The angle-dependent measurements of the ISHE signal[33] were performed in Dev 2 to verify the relation between the direction of the injected spins and the induced charge accumulation in WTe$_2$. The measurements were carried out at different in-plane B field along the tilting angle φ respecting to the x-axis (Fig. 3a). As shown in Fig. 3b, the measured $R_{ISHE}$ decreases with the transverse magnetic (x-direction) component and vanishes when the magnetization is aligned with the y-axis. The sign change of $R_{ISHE}$ is observed between φ=0° and 180° (π) due to switching of the Co magnetization direction and associated reversal of polarization of the spin current. A null $R_{ISHE}$ signal is observed for φ = 90° (π/2) when the magnetization Co is aligned with the y-axis, as the injected spins are parallel to the WTe$_2$ long axis and no ISHE voltage is generated in the measured geometry of WTe$_2$ electrode. The magnitude of the measured ISHE signals $\Delta R_{ISHE}$ as a function of the measurement angle φ is shown in Figure 3c. As expected, charge current **I$_c$** is proportional to **s×I$_s$** (**s** is the spin and **I$_s$** is spin current), the angular dependence of the $\Delta R_{ISHE}$ is expected to vary with cos(φ). Such angle-dependent behaviors essentially show the characteristics of ISHE signal[40]. Figures 3d and 3e show the ISHE signal measured at different spin injection currents, and as expected $\Delta V_{ISHE}$ shows a linear behavior with bias current magnitude. Combining with bias current polarity dependence (Supplementary Fig. S5), we can rules out all the thermal related effects[41,42].

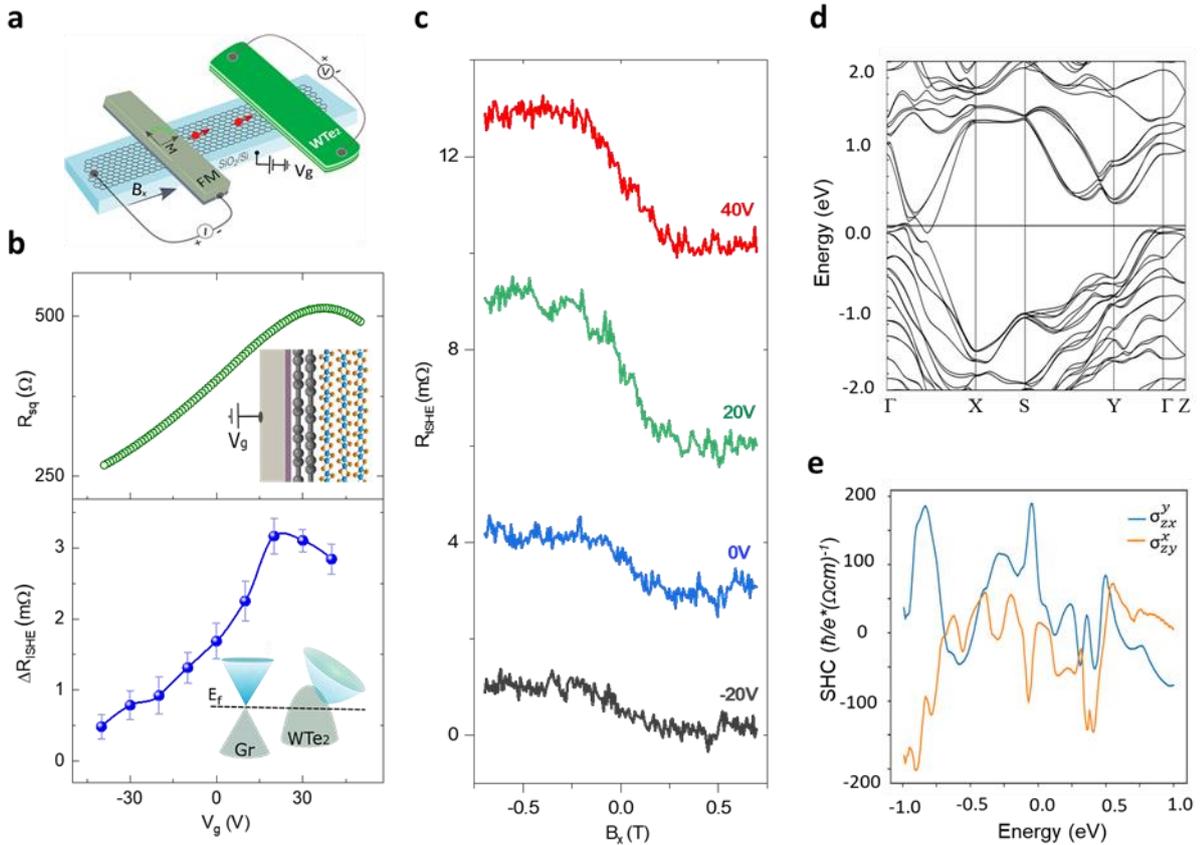

*Figure 4. Gate control of the inverse spin Hall signal in the heterostructure. a. The ISHE geometry for the gate voltage dependent measurements. b. Top panel: The Dirac curve of the few-layer graphene in the*

*heterostructure channel measured in a local four-terminal measurement geometry. Bottom panel: Gate dependence of the ΔR$_{ISHE}$ signal magnitude for Dev 2 measured in the range of V$_g$=-40 to 40 V at room temperature. The line is a guide to the eye. The insets are the schematic and band structure of graphene-WTe$_2$ heterostructure with an applied gate voltage (V$_g$).* **c.** *The measured ISHE resistance R$_{ISHE}$ at different gate voltages with a shift at the y-axis for the sake of clarity.* **d.** *Calculated band structure of bulk WTe2 in Td phase with the Fermi energy is set to zero.* **e.** *Calculated SHE conductivity, $\sigma_{zx}^y$ and $\sigma_{zy}^x$, with respect to the Fermi energy position in WTe2. All the measurements were performed at 300K.*

Gate dependence of ISHE measurement was performed in WTe$_2$-graphene heterostructure (Dev 2) by using the Si/SiO$_2$ as a back gate (Fig. 4a). The gate voltage dependence of the graphene channel resistance across the heterostructure shows the Dirac point at V$_D$= 35 V (Fig. 4b), while WTe$_2$ channel resistance does not show much change due to its semi-metallic character (Fig. S3e). The gate dependence of the graphene-WTe$_2$ interface resistance shows some modulation due to change in the graphene Fermi-energy (Fig. S3d), which may affect the spin absorption efficiency. Figures 4b and 4c show the magnitude of the ISHE signal ΔR$_{ISHE}$ at different gate voltages. Interestingly, we observe a strong increase of the signal magnitude and reaching a maximum as the gate approaches the graphene Dirac point at V$_g$= 30 V. This contrasts with the weak gate dependence of the spin transport signal across the heterostructure (see Supplementary Fig. S8). The gate dependence of spin transport signal (spin valve and Hanle) is known to be strongly dependent on the ferromagnetic tunnel contact resistances[43]. However, in our case, the nonlocal spin transport signal is almost independent of the gate voltage, while the ISHE signals show a strong modulation. This behavior can be explained by considering the spin detector, i.e. WTe$_2$-graphene heterostructure instead of the graphene-ferromagnet impedance mismatch[43]. Part of the enhancement in ΔR$_{ISHE}$ can be due to increased spin absorption by WTe$_2$, i.e. an increase in effective spin current injected vertically at the graphene-WTe$_2$ interface when the graphene resistance increase near the Dirac point[3,4] or (and) the increase of ISHE efficiency with an electric field.

**Discussion**

The SHE signals observed in our experiment can be rationalized by the conventional bulk SHE and the REE in WTe$_2$. In an ideal type-I Weyl semimetal (or an ideal topological insulator), in which the bulk Fermi surface vanishes, the bulk SHE is purely contributed by the topological effect[44]. The bulk –induced spin accumulation on the surface is equivalent to that by the Edelstein effect of the topological Fermi arcs, because of the principle of the bulk-boundary correspondence[45]. However, in a type-II Weyl semimetal, the bulk Fermi states always coexists with the Fermi arcs at the Fermi surface. Therefore, we should consider the SHE to include both Fermi surface states and the bulk states effect. Thus, we perform *ab initio* density-functional theory (DFT) calculations[44,46] (see Methods for details) to evaluate the SHE in WTe$_2$. The calculated electronic band structure and SHE conductivity for WTe$_2$ are shown in Figures 4d and 4e. Here, we consider the SHE that can be quantitatively estimated from the spin Berry curvature of the band structure. The SHE refers to the generation of a spin current $J_j^i = \sigma_{jk}^i E_k$ induced by an electric field $E_k$, where $i, j, k = x, y, z$ (the crystallographic axes of WTe$_2$), and $J_j^i$ represents a spin current along the $j$ direction with a spin polarization along $i$. The SHE conductivity $\sigma_{jk}^i$ characterizes the strength of the SHE. For a charge current along the $ab$ plane and spin current along the $c$ axis, we obtain the corresponding in-plane SHE conductivities in the range of $\sigma_{SH} = 14 - 96\ (\hbar/e)(\Omega \cdot cm)^{-1}$ due to the large crystal anisotropy. Taking the electronic conductivity of WTe$_2$ from our measurement, we estimate the theoretical maximum SHE to be with a spin Hall angle of θ$_{SH}$=$\frac{\sigma_{SH}}{\sigma_{xx}} = 17\%$. The calculated large θ$_{SHE}$ can qualitatively account for the observation of the large SHE signals in our experiments. This is also confirmed by a recent theory paper[47], which suggests a large spin Hall angle in WTe$_2$. Furthermore, as shown in Fig. 4e, the calculated SHE conductivity is also found to be very sensitive and depends on the position of the Fermi energy in WTe$_2$. The calculated $\sigma_{SHE}$ can be considerably tuned by a small change in energy, such as a tiny change in energy at 100 meV below the Fermi energy can cause a change of $\sigma_{SHE}$ by nearly an order of magnitude. However, experimentally we could not tune the surface states of WTe$_2$ to the Fermi level by the application of gate voltage[21,24].

To be noted, one cannot extract spin Hall angle $\theta_{SH}$ and spin diffusion length $\lambda_{WTe2}$ at the same time by fitting the out-of-plane (I)SHE signal[36] or solving the in-plane case equation[38] due to the entanglement of the two parameters. Therefore, we take $\theta_{SH} = 0.013$ from the literature with comparable WTe$_2$ thickness[27]. Consequently, the data fitting results in a spin diffusion length of $\lambda_{WTe2}$=8±0.9nm. To verify the result, a numerical solution to the in-plane ISHE signal was also obtained. Substituting $\lambda_{WTe2}$ into the plot of $\theta_{SH}$-$\lambda_{WTe2}$, we find the $\theta_{SH}$=0.014, which is consistent with the literature (see details in Supplementary Note 2). For estimation of these spin parameters, we used the Dev 3 with monolayer graphene and a narrow and thin WTe$_2$ flake (~ width =1 μm and thickness = 11nm) having a very low WTe$_2$-graphene interface resistance ~ 25 Ω•μm$^2$, which is suitable for the use of the 1D model calculation (shown in Table 1). In contrast, for thicker WTe$_2$ flakes, the WTe$_2$-graphene interface resistance is usually hundreds of Ohms due to their poor van der Waals adhesion. Therefore, an experimental quantification of the SHE parameters in WTe$_2$ by the conventional spin absorption method[38,39] is not possible for Dev 1 and 2. Nevertheless, the calculations suggest that the spin Hall angle $\theta_{SH}$ is larger for the thicker WTe$_2$. This implies a possible presence of a thickness-dependent spin Hall angle and the dominance of SHE in our measurements. More detailed thickness-dependent studies are needed to investigate these effects and disentanglement of the contributions of the SHE and REE in the observed signal[5,48].

**Conclusion and Outlook**

The emergent Weyl semimetal WTe$_2$ is shown here to be a promising material for charge-spin conversion at room temperature due to its unique electronic band structure giving rise to huge spin-orbit coupling and spin-polarized bulk and surface states. Particularly, the strong spin Hall signal in the WTe$_2$-graphene hybrid devices and the gate tunability of the spin absorption process provide a new tool for potential application in future spintronic device architectures. Furthermore, as predicted in theoretical calculations, the spin Hall conductivity can be controlled by using Weyl semimetals with tunable Fermi-level[10] and alloys with tunable-resistivity[37,49]. This will allow achieving systematic control over the charge-spin conversion via electrical and optical means and a better understanding of the Weyl physics. Such measures providing large charge-spin conversion efficiency in Weyl semimetals at room temperature can be used to switch or oscillate the magnetization of nanomagnets with a very low current density. These developments will have a huge potential for emergent spin-orbit induced phenomena and applications in ultralow power magnetic random-access memory and spin logic circuits[29,50].

## Methods

**Device Fabrication** - The exfoliated few layers graphene was mechanically exfoliated onto the n-doped Si substrate with 300 nm SiO$_2$. The CVD graphene was transferred on the substrate Si/SiO$_2$ substrate by wet transfer method and followed by EBL and Ar patterning. The WTe$_2$ flakes were exfoliated on PDMS and dry-transferred on to the graphene flake under a microscope using a home-built micromanipulator transfer stage. The CVD graphene-WTe$_2$ devices were made by exfoliation of WTe$_2$ and dry transfer process inside the glove box. Contacts to graphene and WTe$_2$ were defined by standard electron beam lithography and lift-off process. For the preparation of ferromagnetic tunnel contacts to graphene, a two-step deposition of 0.3 nm of Ti and oxidation process was carried out, followed by a 100 nm of Co deposition. The ferromagnetic tunnel contact (TiO$_2$/Co) resistances on graphene channel were in the range of few kΩs.

**Measurements** - The measurements are performed in a vacuum system equipped with a variable angular rotation facility and with an electromagnet with a magnetic field up to 0.8 Tesla. The electronics used for the measurements are Keithley 6221 current source, Keithley 2182A nanovoltmeter and Keithley 2612B source meter for application of gate voltages.

**Theoretical calculation methods** - We perform ab initio density-functional theory (DFT) calculations and then project the DFT wave functions to atomic-like Wannier functions with the FPLO program[46]. Based on highly symmetric Wannier functions, we construct a tight-binding-type Hamiltonian that can fully reproduce the DFT results. Using the material-specific effective Hamiltonian, we employed the Kubo formula approach[44] to calculate the SHE conductivity.

## Data availability

The data that support the findings of this study are available from the corresponding authors on reasonable request.

## Acknowledgments

The authors at Chalmers University of Technology, Sweden acknowledge financial supports from EU FlagEra project (from Swedish Research Council VR No. 2015-06813), Swedish Research Council VR project grants (No. 2016-03658), EU Graphene Flagship (No. 604391), Graphene center and the AoA Nano program at Chalmers University of Technology. The authors from University of Science and Technology Beijing, Beijing, China, acknowledge financial supports from the National Basic Research Program of China (Grant No. 2015CB921502) and the National Natural Science Foundation of China (Grant Nos. 51731003, 51471029). Bing Zhao would like to thank the financial support from the program of China Scholarships Council (File No.

201706460036) to visit Chalmers. We thank our colleagues at the Quantum Device Physics Laboratory and Nanofabrication Laboratory at Chalmers University of Technology for their support.

**Author information**

**Affiliations**

[1]Beijing Advanced Innovation Center for Materials Genome Engineering, School of Materials Science and Engineering, University of Science and Technology Beijing, Beijing 100083, China

[2]Department of Microtechnology and Nanoscience, Chalmers University of Technology, SE-41296, Göteborg, Sweden

[3]Max Planck Institute for Chemical Physics of Solids, 01187 Dresden, Germany

[4]Department of Condensed Matter Physics, Weizmann Institute of Science, Rehovot 7610001, Israel

[5]Graphene Center, Chalmers University of Technology, SE-41296 Göteborg, Sweden.

**Contributions**

SPD and BZ conceived the idea and designed the experiments. BZ, DK, BK, and SPD fabricated and measured the devices at Chalmers University of Technology. BZ and SPD analyzed, interpreted the experimental data, compiled the figures and wrote the manuscript. DK, BK, AHM, BY, XX, YJ discussed the results and provided feedback on the manuscript. Y.Z., H.F. and B.Y. performed theoretical calculations on the band structure and SHE conductivity. SPD, YJ, BY supervised the experimental and theoretical research. SPD supervised the project.

**Competing interests**

The authors declare no competing financial interests.

**Corresponding authors:**

Correspondence and requests for materials should be addressed to
Saroj P. Dash, Email: saroj.dash@chalmers.se

# Observation of Spin Hall Effect in Weyl Semimetal WTe$_2$ at Room Temperature

Bing Zhao[1,2], Dmitrii Khokhriakov[2], Yang Zhang[3], Huixia Fu[4], Bogdan Karpiak[2], Anamul Md. Hoque[2], Xiaoguang Xu[1], Yong Jiang[1], Binghai Yan[4], Saroj P. Dash[2,5]*

[1]*Beijing Advanced Innovation Center for Materials Genome Engineering, School of Materials Science and Engineering, University of Science and Technology Beijing, Beijing 100083, China*
[2]*Department of Microtechnology and Nanoscience, Chalmers University of Technology, SE-41296, Göteborg, Sweden*
[3]*Max Planck Institute for Chemical Physics of Solids, 01187 Dresden, Germany*
[4]*Department of Condensed Matter Physics, Weizmann Institute of Science, Rehovot 7610001, Israel*
[5]*Graphene Center, Chalmers University of Technology, SE-41296 Göteborg, Sweden.*

Corresponding author: *Saroj P. Dash, Email: saroj.dash@chalmers.se

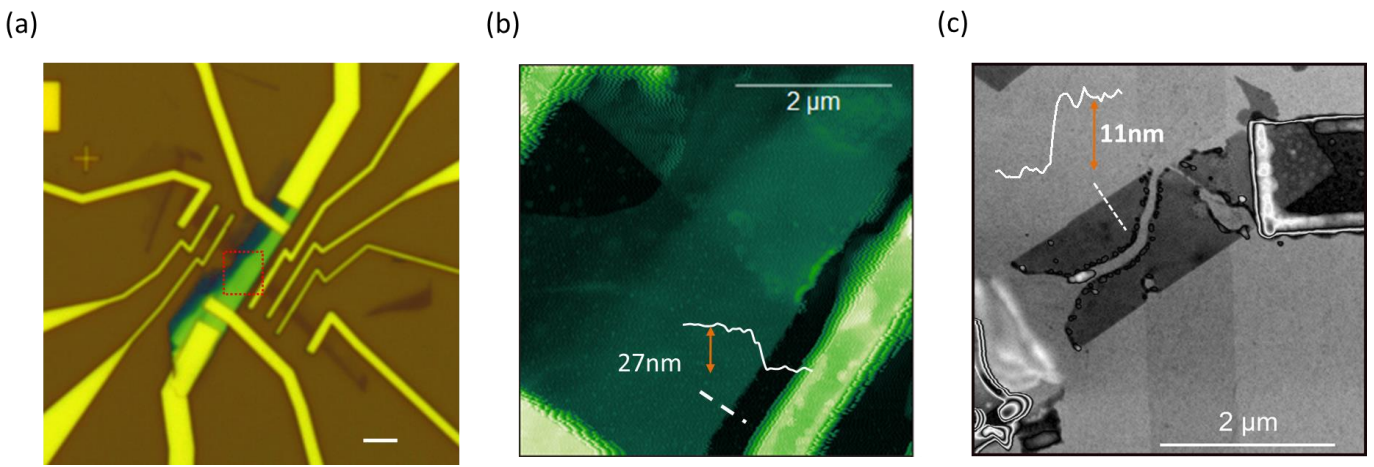

**Figure S1**. **Devices with WTe$_2$-graphene heterostructure.** *(a)* Optical micrograph of Device 1, with WTe$_2$-graphene heterostructure and ferromagnetic (TiO$_2$ 1 nm/Co 60 nm) contacts on graphene for detection and creation of (I)SHE in WTe$_2$. The scale bar is 2 μm. *(b)* Atomic force microscope (AFM) picture of the heterostructure area (red mark in (a)). The inset is the thickness profile of WTe$_2$ on few-layer graphene along the white dash line showing the WTe$_2$ thickness is 27nm in Device 1. *(c)* AFM picture and thickness profile of Dev 3 (after the device burnout) showing WTe$_2$ thickness to be 11 nm on monolayer CVD graphene.

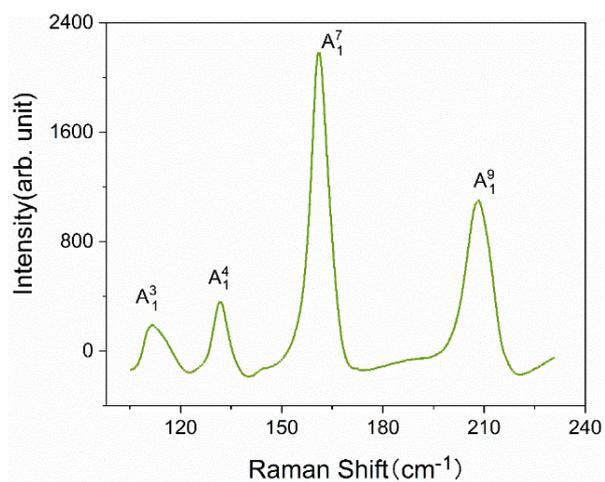

***Figure S2. Raman characterization of $WTe_2$.*** *The Raman spectrum of bulk $WTe_2$ along c- axis using a LASER with a 638 nm wavelength. The characteristic peaks[1] located at about 112, 132, 161, and 208 $cm^{-1}$ corresponding to the $A_1^3$, $A_1^4$, $A_1^7$ and $A_1^9$, showing the $T_d$-phase nature of $WTe_2$.*

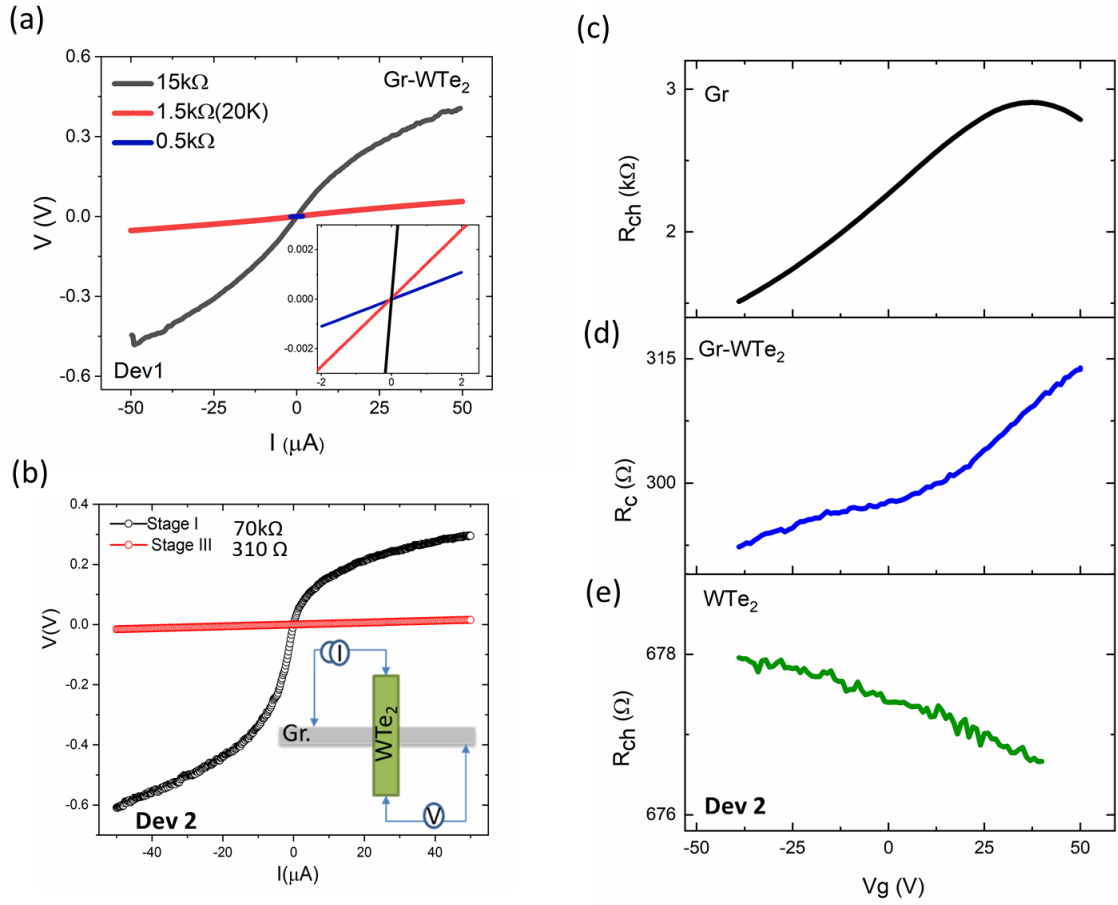

***Figure S3***. ***Electrical characterization of the graphene-WTe₂ heterostructure in Dev 1 and Dev 2***. *(**a**) The current-voltage (IV) characteristics of the graphene-WTe₂ interface resistance of Dev 1 at different stages measured in a four-terminal geometry. The inset is a zoom-in part around zero bias. The interface resistance of graphene-WTe₂ is about 15 kΩ at stage II, which is reduced to 500 Ω at stage III. (**b**) Current-voltage (IV) curve of the WTe₂-graphene interface contact at different stages of resistances, stage I= 70 kΩ and stage III= 310 Ω at room temperature. The inset shows the schematic of the four-terminal geometry used for the IV measurements. Ultimately, all the devices stabilized at a low contact resistance of 300-500 Ω, which is suitable for (I)SHE measurements. (**c-e**) Gate dependence of graphene (Gr) channel resistance, WTe₂-graphene contact resistance and WTe₂ channel resistance in Dev 2, respectively.*

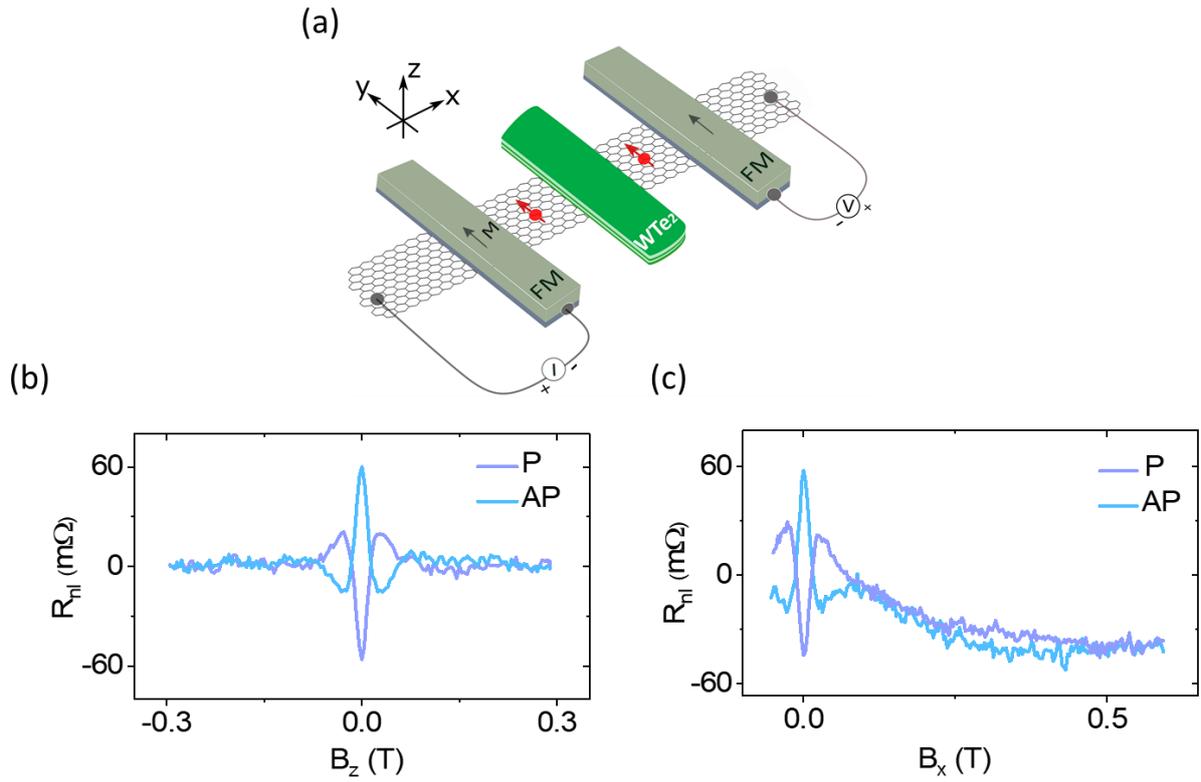

*Figure S4. **Spin transport and precession across the graphene-WTe$_2$ heterostructure before the spin absorption in Dev 2.** (a) Schematic of the non-local Hanle measurement geometry in the WTe$_2$-graphene heterostructure device. (b, c) Hanle spin precession signal for different magnetic field sweep directions B$_z$ and B$_x$ for both the parallel (P) and anti-parallel (AP) alignment of ferromagnetic Co electrodes in Dev 2 at room temperature in stage1 (before the spin absorption). A linear background is removed from the measured spin signals.*

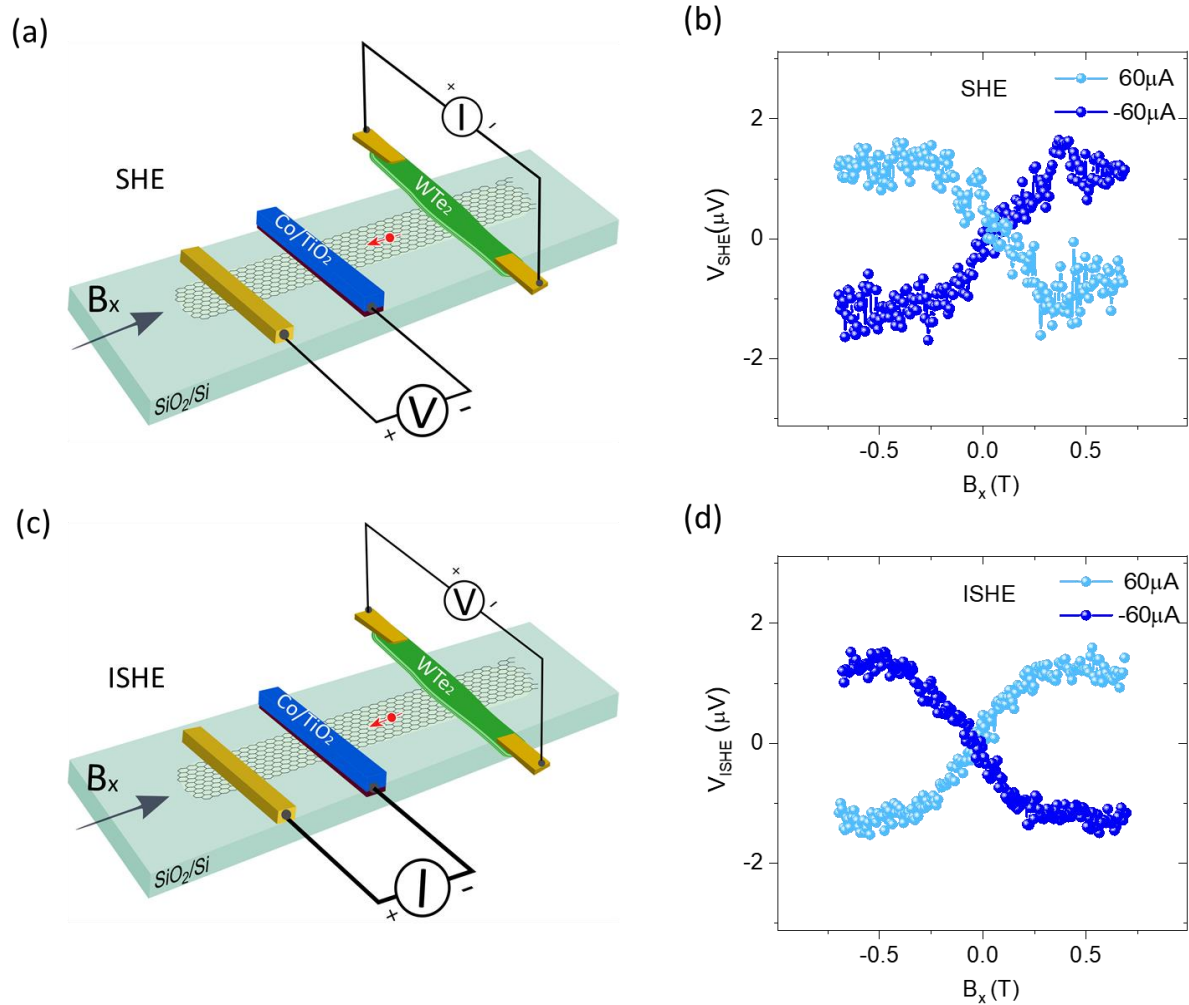

*Figure S5. **Charge-to-spin conversion (SHE) and inverse charge-to-spin conversion (ISHE) effect in graphene-WTe$_2$ heterostructures with different bias polarity for Dev 1**. **(a, b)** SHE measurement geometry and data with an applied bias current of +60µA and -60µA along with the WTe$_2$ flake. The spin current induced at the WTe$_2$-graphene interface diffuse towards the Co/TiO$_2$ detector contact. **(c, d)** ISHE measurement geometry and the measured signal, where the spin current is injected from the Co/TiO$_2$ contact with +60µA and -60µA bias currents. The injected spin current diffuse towards the WTe$_2$-graphene heterostructure and produce the ISHE signal. For the sake of clarity, the linear background is subtracted from the (I)SHE signals.*

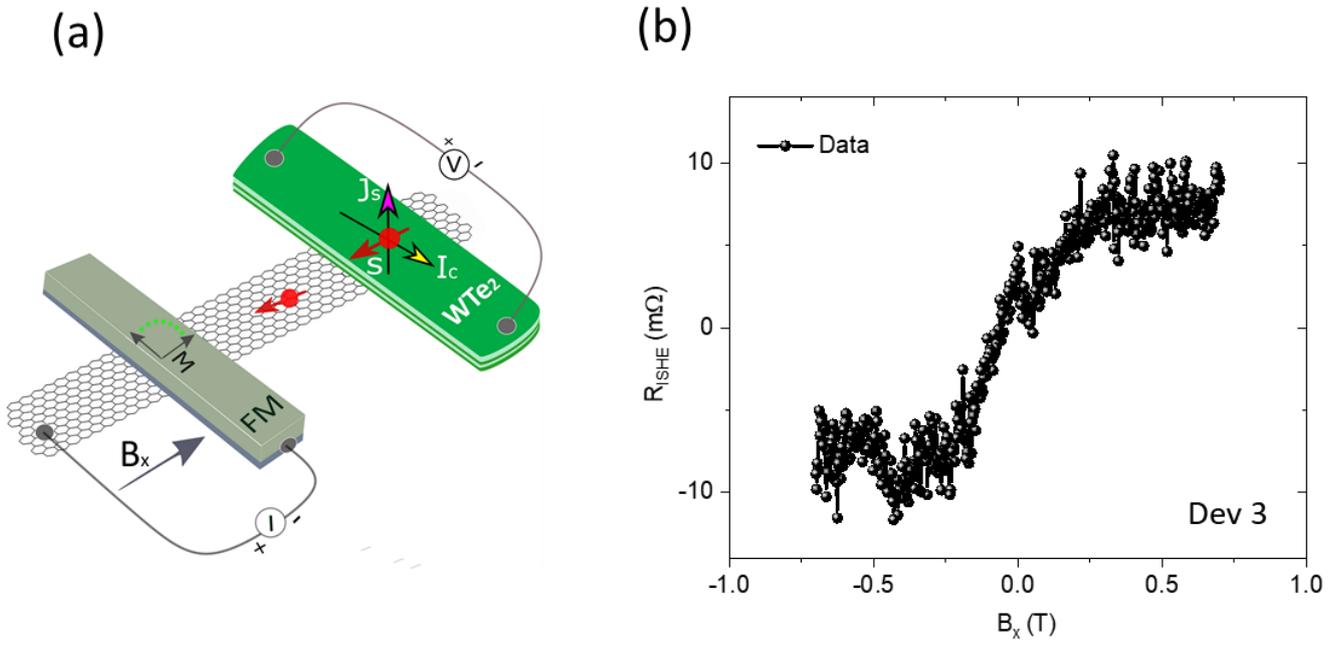

*Figure S6. **Inverse spin Hall effect (ISHE) in monolayer graphene-WTe$_2$ heterostructure in Dev 3. (a, b)** ISHE measurement geometry and measured signal with sweep of magnetic field $B_x$ in Dev 3 with spin injection bias current of 100µA and $V_g$=-60V, at room temperature. The Dev 3 consists of monolayer CVD graphene and a narrow and thin WTe$_2$ flake (~ width =1 µm and thickness = 11nm) heterostructure with interface resistance 25 Ωµm$^2$. A linear background is subtracted from the signal.*

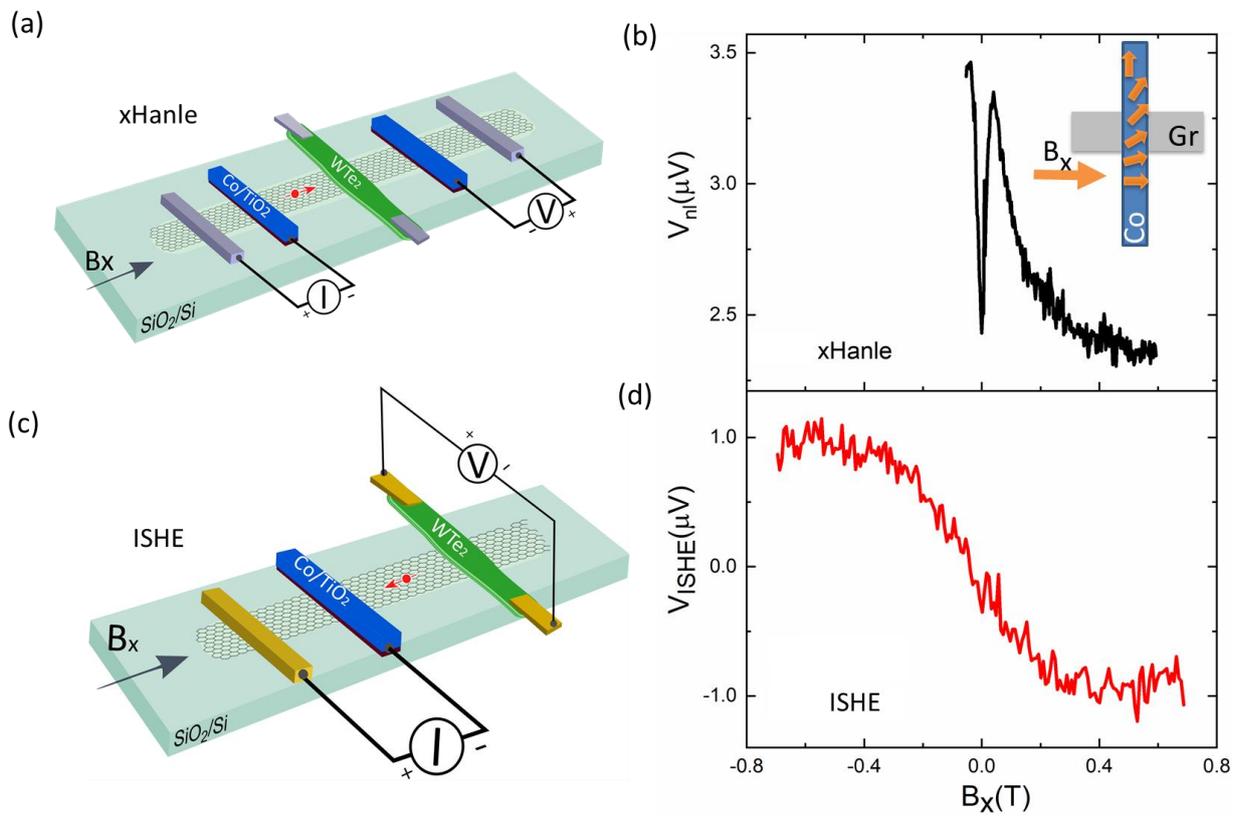

*Figure S7. **Estimation of magnetization saturation field from Hanle measurements.** **(a, b)** Hanle spin precession measurement geometry and data for the graphene-WTe$_2$ heterostructure channel with magnetic field sweep in the direction of $B_x$ at room temperature for Dev2 at stage-I. Inset in (b) is the schematic for the rotation of the Co magnetization when sweeping a B field in the x-direction. **(c, d)** Inverse SHE measurement geometry and data with the use of same Co as spin injector contact in Dev 2. Both measurements were applied with -60μA.*

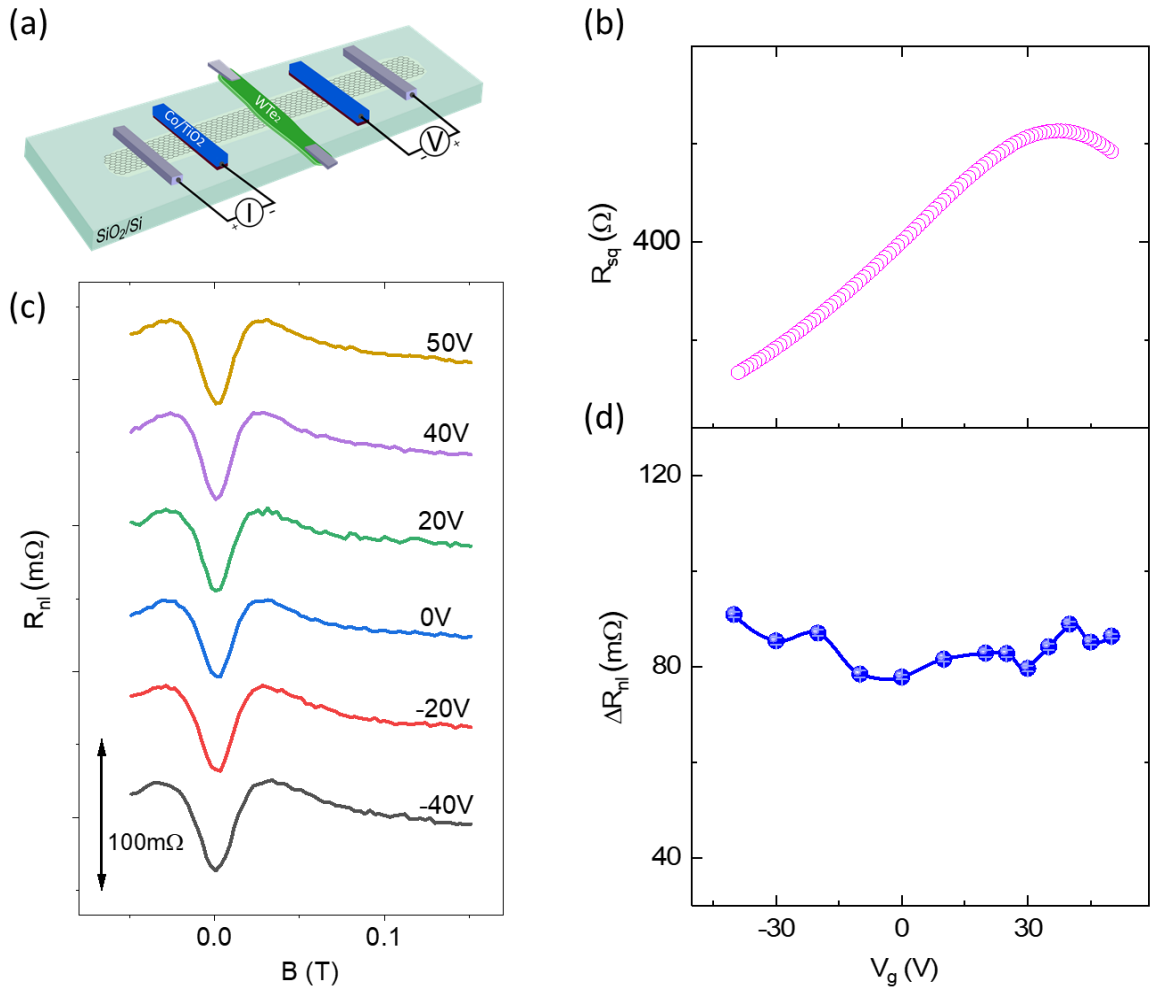

***Figure S8***. ***Gate dependence of nonlocal spin signal across the graphene-WTe$_2$ heterostructure of Dev 2 in stage-I, before spin absorption***. *(a) Schematics of the nonlocal spin transport measurement geometry.  (b, c, d) Gate dependence of the graphene channel square resistance, Hanle spin precession signal, and the magnitude of the spin signal across the WTe$_2$-Gr heterostructure at stage I (before spin absorption) in Dev 2 at room temperature, respectively. The error bars are within the data points.*

# Note 1. Spin absorption effect at the WTe$_2$-graphene interface

The spin current in the graphene channel can be absorbed by WTe$_2$ due to its high SOC[2], highly depending on the contact resistance of the graphene-WTe$_2$ interface[3,4]. We observed that the van der Waals interface contact resistance of graphene-WTe$_2$ usually evolved during the measurements (Fig. S3a and S3b) in the exfoliated few-layer graphene and 20-30 nm WTe$_2$ devices (Dev 1 and Dev 2). Three distinctly different resistance states of the graphene-WTe$_2$ interface were observed at room temperature. As shown in Fig. S3b, stage-I shows a tunneling type conduction, with zero-bias contact resistance (ZCR) of about 70 kΩ. Stage-II is an intermediate state with ZCR around 3kΩ (not shown in the figure). Ultimately the stage-III show a lower interface with a very stable ZCR of 310 Ω. Similar behaviors are observed in both Dev 1 and Dev 2. Due to the decrease of the resistance and the ultimate formation of low stable interface resistance between graphene and WTe$_2$, the oxidation of WTe$_2$ at the interface can be ruled out. Such a contact behaviour is believed to be caused by a gap or air bubble at the interface introduced in the WTe$_2$ flake dry transfer process. During the fabrication, the use of a vacuum treatment of the heterostructure before spin coating with e-beam resist, resulted in a faster evolution of the interface resistance. Moreover, the graphene channel covered by WTe$_2$ is found to be very stable with an almost unchanged Dirac Point and other parameters (see Table 1), which shows a harmonious van der Waals bonding nature of graphene and WTe$_2$. With the evolution of graphene-WTe$_2$ interface resistance, the spin current in the graphene channel gets increasingly absorbed by the WTe$_2$ at the interface. Direct evidence for this is the obvious decrease of the spin valve signal by 81.3% from 106 mΩ to 20 mΩ as the contact resistance changes from stage I=70 kΩ to stage III= 310 Ω (Fig. S9c). Similar behavior is also observed in Dev 1 (about a 50% decrease, see Fig. S9b). Moreover, the TiO$_2$/Co contact resistances and graphene channel resistance at stage-I and -III remain comparable. Therefore, we can conclude that the strong spin absorption effect is the origin for the reduction of the spin valve signal from the state I to stage III[3,4], which allowed us to measure the ISHE signals in our devices. As expected, no (I)SHE signals could be measured for higher graphene-WTe$_2$ contact resistances in stage I, because of very-low spin absorption at the interface.

However, in the Dev 3, the heterostructure of CVD monolayer graphene and 11 nm WTe$_2$ devices, a much lower interface resistance 25 Ωμm$^2$ is obtained (see Table 1 for Dev 3), which guarantees the transparent interface between WTe$_2$ and graphene. In this device, no spin valve signal could be observed across the heterostructure channel (Fig. S9d), which suggests about 100% spin absorption at the WTe$_2$-graphene interface. To verify the spin injection and detection of the ferromagnetic contacts of Co/TiO$_2$ in the Dev 3, the spin valve and Hanle measurement of the only area graphene channel outside the WTe$_2$ were also performed. The standard spin valve and Hanle signal can be observed. Furthermore, the use of a narrow WTe$_2$ (1μm), less than spin diffusion length in graphene, makes sure that the 1D model can be used in the analysis.

# Table 1

**Parameters of the devices.** $L_{ch,Gr}$, $w_{Gr}$, $R_{sq,Gr}$, $D_s$, $\tau_s$, $\lambda_{Gr}$, are the channel length, width, square resistance, spin diffusion constant, spin lifetime and spin diffusion length of graphene, respectively. $t_{WTe2}$, $w_{WTe2}$, $\rho_{WTe2}$ are the thickness, width, resistivity of $WTe_2$, respectively. $L_{SH}$ is the channel length of (I)SHE measurement. $R_{c,WTe2-Gr}$ is the contact resistance of the $WTe_2$-Gr heterostructure. $R_cA$ is the product of the heterostructure contact resistance and area. Dev 1 and Dev 2 uses few layers of exfoliated graphene, where as Dev 3 is made up of monolayer CVD graphene.

| | | $L_{ch,Gr}$ (µm) | $D_s$ (m²/s) | $\tau_s$ (ps) | $\lambda_{Gr}$ (µm) | $R_{c,WTe2-Gr}$ (kΩ) | $w_{Gr}$ (µm) | $L_{SH}$ (µm) | $t_{WTe2}$ (nm) | $w_{WTe2}$ (µm) | $R_cA$ (kΩµm²) | $R_{sq,Gr}$ (kΩ) | $\rho_{WTe2}$ (µΩm) |
|---|---|---|---|---|---|---|---|---|---|---|---|---|---|
| Dev 1 | stage I | 5.1 | 0.0141 | 316 | 2.11 | 50 | 1.35 | 2.6 | 27 | 3 | 203 | 0.423 | 8.9 |
| | stage III | | 0.0188 | 202 | 1.95 | 0.5 | | | | | 2.03 | | |
| Dev 2 | stage I | 6.9 | 0.0154 | 326 | 2.24 | 70 | 1.3 | 3.5 | 20 | 5 | 440 | 0.496 | 8.9 |
| | stage III | | 0.0178 | 299 | 2.31 | 0.31 | | | | | 1.95 | | |
| Dev 3 | $V_g=-60V$ | 6.9 | 0.03011 | 185 | 2.4 | 0.025 | 1 | 3.5 | 11 | 1 | 0.025 | 0.546 | 9.04 |

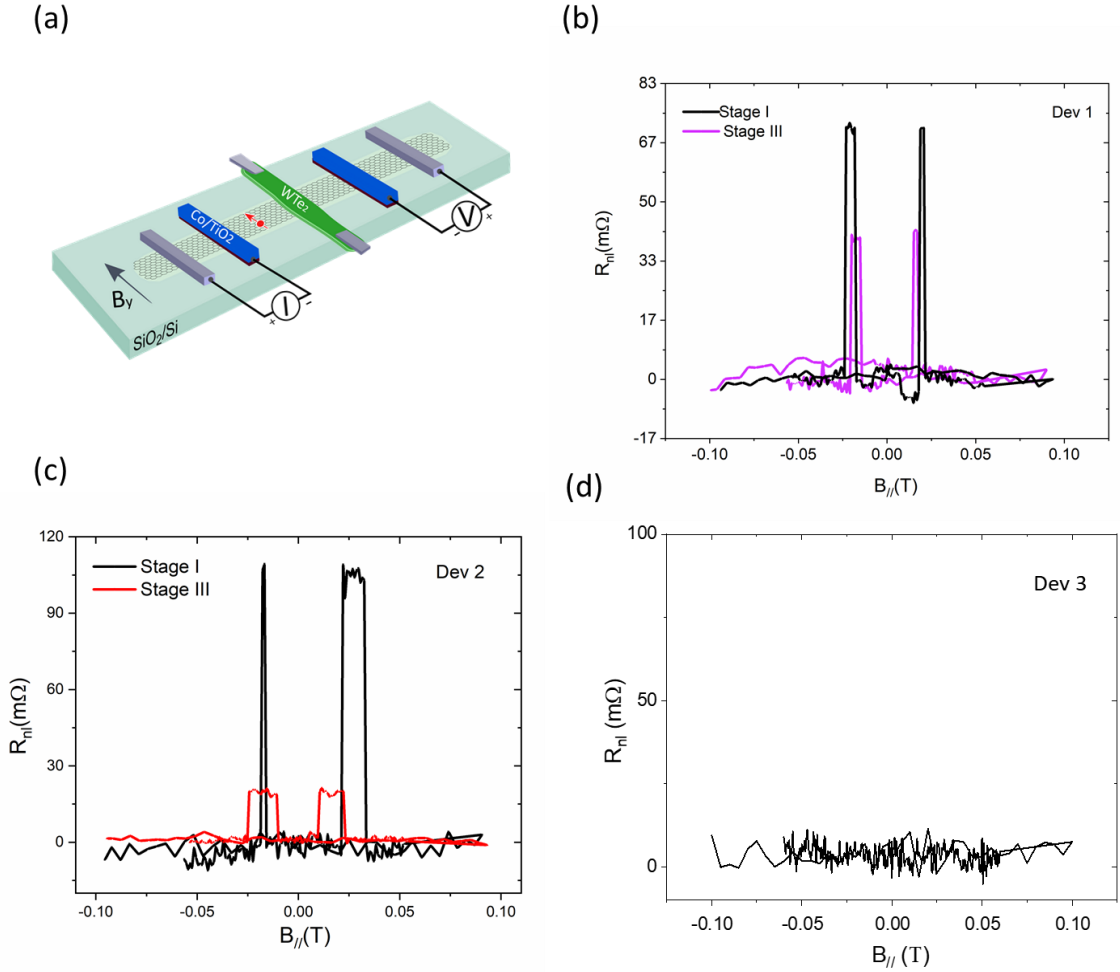

***Figure S9**. **Spin absorption at the graphene-WTe2 interface.** **(a)** The spin-valve measurement geometry and the signal measured in the graphene channel in the heterostructure with WTe2. **(b,c)** Due to the change of WTe2-graphene interface resistances from stages-I to –III, we observed the large reduction of spin signal $R_{nl}$ in Dev 1 and 2. **(d)** No spin signal was observed in the noise level across the heterostructure in Dev 3. To be noted, all the spin valve signals are measured at negative bias currents and the $R_{nl}=V_{nl}/I$, where I is the absolute value of the bias current magnitude.*

# Note 2. Estimation of the Spin Hall Angle and Spin Diffusion Length in WTe$_2$

The calculation of shunting factor x is very crucial[2,5] for (I)SHE measurement and for the verification of the possible SHE origins. The shunting factor is defined as the relative value under the assumption that all the charge current flows within WTe$_2$, while (1-x) means the corresponding current shunted by graphene. We use the 3D COMSOL AC/DC module to calculate current distribution in our WTe$_2$-graphene heterostructure with all the parameters taken from the devices. Our calculation result shows that the shunting effect is highly suppressed due to the relatively higher interface resistance of WTe$_2$-graphene and the comparable conductivity of the two materials. Therefore, the shunting factor x≈1. Therefore, all the shunting related origins can be ruled out, such as proximity-induced SHE in graphene and proximity induced Rashba-Edelstein effect in graphene[6,7].

To estimate the spin Hall angle θ$_{SH}$ and spin diffusion length λ$_{WTe2}$ in WTe$_2$, usually one should obtain the spin diffusion length from spin absorption experiments[2,8], in which one can estimate the spin diffusion length from the reduction of spin signal due to the presence of WTe$_2$ in the graphene channel. However, experimentally it is challenging to compare different graphene spintronic devices with and without WTe$_2$, as FM contact spin polarization, tunnel resistances and graphene spin transport parameters and channel doping level are not always comparable in the device with and without the WTe$_2$ layer. These factors can affect the magnitude of the spin signals and spin lifetime in the channels. So, it is not reliable to obtain the λ$_{WTe2}$ by the spin absorption experiments. Therefore, here we use both the in-plane ISHE[2,8] and out-of-plane ISHE[9] method in the same device (Dev 3) for estimation of spin parameters. By comparing the results from these two methods, we can check the self-consistency and reliability of the calculations.

**Out-of-plane ISHE -** For the out-of-plane ISHE, we adopt the formula as below[9],

$$R_{ISHE}(B_z) = \frac{p_i \theta_{SH} \rho_M L_{SH} \lambda_{WTe2}}{w_M t_M} \left[\frac{1-\exp(-\frac{t_M}{\lambda_{WTe2}})}{1+\exp(-\frac{t_M}{\lambda_{WTe2}})}\right] \int_0^\infty \frac{\exp(-\frac{L_{SH}^2}{4D_s t})}{t\sqrt{4\pi D_s t}} \sin(\omega_0 B_z t)\exp(-\frac{t}{\tau_s})dt, \quad\quad S(1)$$

Where $\rho_M$, $w_M$, $t_M$, $\theta_{SH}$ and $\lambda_{WTe2}$ are resistivity, width, thickness, spin Hall angle and spin diffusion length of WTe$_2$, respectively. $L_{SH}$ is graphene channel length, $p_i$=0.32 and D$_s$=0.032m$^2$/s are the effective spin polarization of Co/TiO$_2$ and spin diffusion constant which are extracted from the standard Hanle fitting of the spin valve nearby the WTe$_2$. Here we assume the D$_s$ does not change in all the graphene area. $\tau_s$ is the spin diffusion time in graphene. $\omega_0 = g\mu_B/\hbar$ is the Larmor precession frequency, where g=2 is Lande g-factor, $\mu_B$ and $\hbar$ are the Bohr magneton and reduced Planck constant. If the θ$_{SH}$=0.013 is taken from the literature[10], one can extract that the λ$_{WTe2}$ =8±0.9nm and $\tau_s$=185±9ps (main text Fig.2c) from the Eq. S1 fitting. The spin resistance $R_{s,WTe2} = \frac{\rho_M \lambda_{WTe2}}{w_M w_{gr} tanh(t_M/\lambda_{WTe2})} \approx 80 m\Omega \ll R_{s,gr}=\frac{R_{sq,gr}\lambda_{gr}}{w_{gr}} \approx 1.2k\Omega$ which is vital to guarantee the accurate extraction for the shunting effect can be highly suppressed[9].

**In-plane ISHE -** In order to quantitatively understand the in-plane ISHE data from the same device, we use a model based on one-dimensional spin diffusion equations. The detected ΔR$_{ISHE}$ can be expressed as[2,11,12]:

$$\Delta R_{ISHE} = \frac{2\theta_{SH}\rho_M x}{w_M}\left(\frac{\hat{I}_s}{I_c}\right) \quad\quad S(2)$$

Where x is the shunting factor. The effective spin current $\hat{I}_s$ injected vertically from graphene into the WTe$_2$ results in a charge current of $I_c$. Where,

$$\frac{\hat{I}_s}{I_c} \equiv \frac{\int_0^{t_M} I_s(z)dz}{I_c \lambda_{WTe2}} = \frac{\lambda_{WTe2}}{t_M} \frac{(1-e^{-\frac{t_M}{\lambda_{WTe2}}})^2}{1-e^{-2\frac{t_M}{\lambda_{WTe2}}}} \frac{I_s(z=0)}{I_c}$$

$$= \frac{\lambda_M}{t_M} \frac{(1-e^{-\frac{t_M}{\lambda_{WTe2}}})^2}{1-e^{-2\frac{t_M}{\lambda_{WTe2}}}} \frac{2p_i Q_{IF1}[(2Q_{IF2}+1)(1-Q_{IM})e^{\frac{-L_{SH}}{\lambda_{gr}}} - (1+Q_{IM})e^{\frac{-3L_{SH}}{2\lambda_{gr}}}]}{(2Q_{IF1}+1)(2Q_{IF2}+1)(1+Q_{IM}) - (2Q_{IF1}+1)(1+Q_{IM})e^{\frac{-3L_{SHE}}{2\lambda_{gr}}} - (2Q_{IF2}+1)(1-Q_{IM})e^{\frac{-L_{SH}}{\lambda_{gr}}} - (Q_{IM}-1)e^{\frac{-2L_{SH}}{\lambda_{gr}}}} \quad (S3)$$

Where $R_{gr} = \frac{R_{gr}^\blacksquare \lambda_{gr}}{w_{gr}}$, $Q_{IFi} = \frac{1}{1-p^2}\frac{R_{ci}}{R_{gr}}$, with i=1, 2, $R_{ci}$ corresponding to injector and detector contact resistance, respectively. $Q_{IM} = \frac{w_{gr} R_{cM}}{R_{gr}^\blacksquare \lambda_{gr}}$, $R_{cM}$, $w_{gr}$, $R_{gr}^\blacksquare$ are WTe2-graphene interface resistance, graphene channel width and square resistance, respectively. By analyzing the Eq. S2 and substituting all the device parameters from Dev 3 (Table 1 and Fig. S6), one can obtain the plot of the relation between the spin Hall angle $\theta_{(I)SHE}$ and spin diffusion length $\lambda_{WTe2}$ (Fig.S10). As expected, by substituting $\lambda_{WTe2}$=8±0.9nm from the out-of-plane ISHE result, one obtains $\theta_{SH}$=0.014, which is comparable to the value in the literature[10] and proves the self-consistency in the calculations.

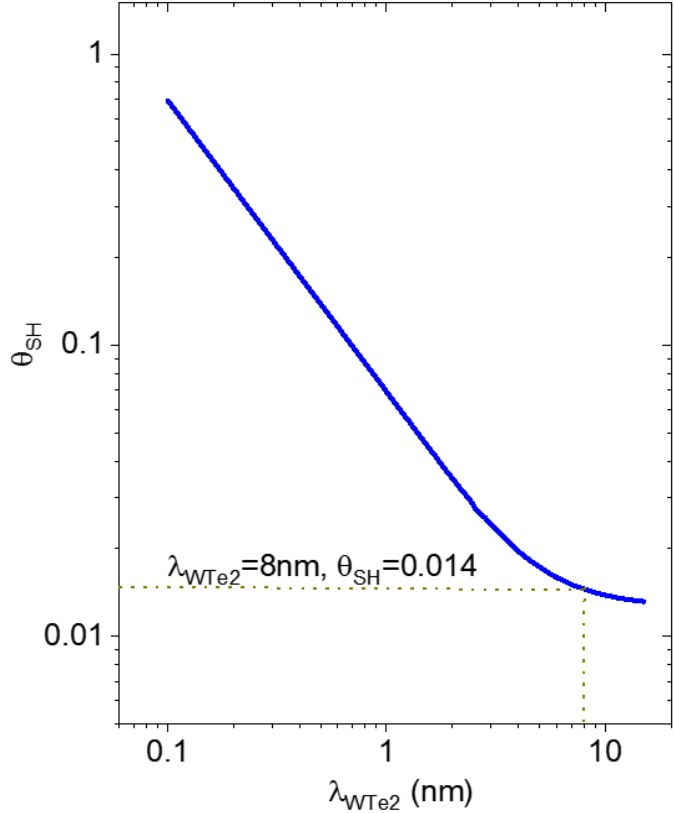

**Figure S10**. *Estimation of spin Hall angle and spin diffusion length of WTe2. Calculated spin parameters ($\lambda_{WTe2}, \theta_{SH}$) in WTe2 using a numerical solution to the Eq. S2 and parameters from Dev 3. The plot is the relation between spin Hall angle $\theta_{SH}$ and spin diffusion length $\lambda_{WTe2}$. The dashed lines show one possible solution.*